\documentclass[pra,aps,showpacs,twocolumn,longbibliography]{revtex4}
\usepackage{amssymb}
\usepackage{amsmath}
\usepackage{mathtools}
\usepackage[dvipdfm]{hyperref}
\usepackage{hyperref}
\usepackage{stmaryrd}
\usepackage{float}

\allowdisplaybreaks
\begin{document}

\title{Quantifying the imaginarity of quantum states via Tsallis relative entropy}
\author{Jianwei Xu}
\email{xxujianwei@nwafu.edu.cn}
\affiliation{College of Science, Northwest A$\&$F University, Yangling, Shaanxi 712100,
China}

\begin{abstract}
It is a fundamental question that why quantum mechanics uses complex numbers
instead of only real numbers. To address this topic, recently, a rigorous
resource theory for the imaginarity of quantum states were established, and
several imaginarity measures were proposed. In this work, we propose a new
imaginarity measure based on the Tsallis relative entropy. This imaginarity
measure has explicit expression, and also, it is computable for bosonic Gaussian states.
\end{abstract}

\maketitle
\date{\today }


\section{Introduction}

A postulate of quantum mechanics says that \cite{Nielsen-2010-quantum} a
quantum system is described by a complex Hilbert space. From the inception
of quantum mechanics, there is a debate that whether complex numbers are
used only for the mathematical convenience or are necessary in describing
quantum systems. To address this topic, recently, a rigorous imaginarity
theory of quantum states were established
\cite{JPA-Gour-2018,PRL-Guo-2021,PRA-Guo-2021}. This imaginarity theory can
be viewed as a quantum resource theory \cite{IJMPB-Horodechi-2013,RMP-Gour-2019}.

We consider a quantum system associated with the $d$-dimensional complex
Hilbert space $H.$ In imaginarity theory, an orthonormal basis $\{|j\rangle
\}_{j=1}^{d}$ of $H$ is fixed, i.e., imaginarity theory is basis dependent.
A quantum state is represented by a density operator $\rho .$ We express $%
\rho $ in basis $\{|j\rangle \}_{j=1}^{d}$ as
\begin{eqnarray}
\rho =\sum_{j,k=1}^{d}\rho _{jk}|j\rangle \langle k|, \label{eq1-1}
\end{eqnarray}%
where $\rho _{jk}=\langle j|\rho |k\rangle .$ $\rho $ is called a real state
if $\rho _{jk}=\langle j|\rho |k\rangle $ is a real number for any $j$ and $%
k.$ We decompose $\rho $ as
\begin{eqnarray}
\rho =\text{Re}\rho +i\text{Im}\rho , \label{eq1-2}
\end{eqnarray}%
where $i=\sqrt{-1},$ $\text{Re}\rho =\sum_{j,k=1}^{d}(\text{Re}\rho
_{jk})|j\rangle \langle k|$ is the real part of $\rho ,$ $\text{Im}\rho
=\sum_{j,k=1}^{d}(\text{Im}\rho _{jk})|j\rangle \langle k|$ is the imaginary
part of $\rho ,$ $\text{Re}\rho _{jk}$ and $\text{Im}\rho _{jk}$ are the
real part and imaginary part of $\rho _{jk}.$ We see that $\rho $ is a real
state if and only if $\text{Im}\rho =0.$ Since $\rho =\rho ^{\dagger },$
then $\text{Re}\rho $ is symmetric and $\text{Im}\rho $ is antisymmetric,
that is, $(\text{Re}\rho )^{T}=\text{Re}\rho $ and $(\text{Im}\rho )^{T}=-%
\text{Im}\rho .$ We use $()^{\ast },$ $()^{T}$ and $()^{\dagger }$ to denote
the (complex) conjugate, transpose and conjugate transpose of the matrix $%
(), $ respectively.

A quantum operation $\phi $ can be represented by a set of Kraus operators $%
\phi =\{K_{l}\}_{l}$ satisfying $\sum_{l}K_{l}^{\dagger }K_{l}\preceq I$
\cite{Nielsen-2010-quantum}. Here $I$ is the identity operator, $%
\sum_{l}K_{l}^{\dagger }K_{l}\preceq I$ means $I-\sum_{l}K_{l}^{\dagger
}K_{l}\succeq 0,$ that is, $I-\sum_{l}K_{l}^{\dagger }K_{l}$ is positive
semidefinite. The quantum operation $\phi =\{K_{l}\}_{l}$ is called a
quantum channel if $\sum_{l}K_{l}^{\dagger }K_{l}=I.$ The quantum operation $%
\phi =\{K_{l}\}_{l}$ performs the state $\rho $ into the (unnecessarily normalized) state $\phi (\rho
)=\sum_{l}K_{l}\rho K_{l}^{\dagger }.$ In imaginarity theory, a quantum
operation $\phi =\{K_{l}\}_{l}$ is called a real operation if $K_{l}$ is a
real matrix in the basis $\{|j\rangle \}_{j=1}^{d}$ for any $l.$ We see that
$\phi (\rho )$ is a real (unnecessarily normalized) state if $\rho $ is a real state and $\phi
=\{K_{l}\}_{l}$ is a real operation.

With the definitions of real states and real operations, some conditions are
proposed for an imaginarity measure. A real-valued functional $M$ on quantum
states is called an imaginarity measure if $M$ satisfies the following
conditions (M1) to (M4) \cite{JPA-Gour-2018,PRL-Guo-2021,PRA-Guo-2021}.

(M1). Faithfulness: $M(\rho )\geq 0$ and $M(\rho )=0$ if and only if $\rho $
is real.

(M2). Monotonicity: $M[\phi (\rho )]\leq M(\rho )$ if $\phi $ is a real
channel.

(M3). Probabilistic monotonicity: $\sum_{l}$tr$(K_{l}\rho K_{l}^{T})M[%
\frac{K_{l}\rho K_{l}^{T}}{\text{tr}(K_{l}\rho K_{l}^{T})}]\leq M(\rho )$ if
$\phi =\{K_{l}\}_{l}$ is a real channel.

(M4). Convexity: $M(\sum_{j}p_{j}\rho _{j})\leq \sum_{j}p_{j}M(\rho _{j})$
for any probability distribution $\{p_{j}\}_{j}$ and states $\{\rho
_{j}\}_{j}.$

Note that (M3) and (M4) together imply (M2).
In Ref. \cite{QIP-2021-Li}, the authors considered the condition (M5) below.

(M5). Additivity for direct sum states.
\begin{eqnarray}
M[p\rho _{1}\oplus (1-p)\rho _{2}]=pM(\rho _{1})+(1-p)M(\rho _{2}), \label{eq1-3}
\end{eqnarray}
where $p\in (0,1),$ $\{\rho _{1},\rho _{2}\}$ are any quantum states.

It is shown \cite{QIP-2021-Li} that (M3) and (M4) are equivalent to (M2)
and (M5). Consequently, to check that whether a real-valued functional on
quantum states is a imaginarity measure, we can check (M1), (M3) and (M4),
or we can check (M1), (M2) and (M5). This equivalence is similar to the case
of quantum coherence \cite{PRA-2016-Tong,PRA-2019-Tong,PRA-2020-Xu}.

Several imaginarity measures have been found, such as the trace norm of
imaginarity \cite{JPA-Gour-2018,PRL-Guo-2021}, denoted by $M_{\text{tr}%
}(\rho );$ the relative entropy of imaginarity \cite{QIP-2021-Li}, denoted
by $M_{\text{r}}(\rho )$; and $M_{\text{F}}(\rho )$ based on the fidelity
\cite{PRA-Guo-2021,NJP-2022-Streltsov}. They are
defined as
\begin{eqnarray}
M_{\text{tr}}(\rho ) &=&||\rho -\rho ^{\ast }||_{\text{tr}},  \label{eq1-4} \\
M_{\text{r}}(\rho ) &=&S(\text{Re}\rho )-S(\rho ),  \label{eq1-5} \\
M_{\text{F}}(\rho ) &=&1-F(\rho ,\rho ^{\ast }),  \label{eq1-6}
\end{eqnarray}
where $F(\rho ,\sigma )=1-$tr$\sqrt{\sqrt{\rho }\sigma \sqrt{\rho }}$ is the
fidelity of states $\rho $ and $\sigma .$

Imaginarity theory is in active research in these years \cite{PRL-Guo-2021,Nature-2021-Acin,PRAP-2021-Guo,PRR-2021-Zhu,PRL-2022-Li,NJP-2022-Streltsov,PRL-2023-Streltsov,arXiv-2023-Xu}. State transformations under real operations have been extensively studied
\cite{NJP-2022-Streltsov}, the imaginarity of
bosonic Gaussian states has been discussed \cite{arXiv-2023-Xu}. Some results of imaginary theory were experimentally
investigated \cite{PRL-Guo-2021,Nature-2021-Acin,PRAP-2021-Guo,PRR-2021-Zhu,PRL-2022-Li}.

In this work, we propose a new imaginarity measure based on the Tsallis
relative entropy, we denote this imaginarity measure by $M_{T,\mu }(\rho ).$
We will show that $M_{T,\mu }(\rho )$ possesses explicit expression for
finite-dimensional quantum states, and remarkably, $M_{T,\mu }(\rho )$ is
also computable for general bosonic Gaussian states in terms of means and
covariances of Gaussian states. This paper is organized as follows. In
section II, we provide some properties about the real operations. In section III, we link $M_{T,\mu }(\rho )$ to some other quantities and then provide some interpretations for $M_{T,\mu }(\rho )$. In section V, we find out the expression of $M_{T,\mu }(\rho )$
for general bosonic Gaussian states in terms of the means and covariances of
Gaussian states. Section VI is a brief summary.

\section{Properties of real operations}

In this section, we provide some properties of real operations.

\emph{Proposition 1.} For any state $\rho $ and any real operation $\phi
=\{K_{l}\}_{l},$ it holds that
\begin{eqnarray}
\phi (\rho ^{\ast })=[\phi (\rho )]^{\ast }.  \label{eq2-1}
\end{eqnarray}

\emph{Proof.} Using Eq. (\ref{eq1-2}) and the fact that $\{K_{l}\}_{l}$ are all real
matrices in basis $\{|j\rangle \}_{j=1}^{d},$ we have
\begin{eqnarray}
\phi (\rho ^{\ast }) &=&\sum_{l}K_{l}(\text{Re}\rho -i\text{Im}\rho
)K_{l}^{T} \nonumber \\
&=&\sum_{l}K_{l}(\text{Re}\rho )K_{l}^{T}-i\sum_{l}K_{l}(\text{Im}\rho
)K_{l}^{T}  \nonumber  \\
&=&[\sum_{l}K_{l}(\text{Re}\rho )K_{l}^{T}+i\sum_{l}K_{l}(\text{Im}\rho
)K_{l}^{T}]^{\ast }  \nonumber \\
&=&[\phi (\rho )]^{\ast }.   \nonumber
\end{eqnarray}
$\hfill\blacksquare$

Proposition 1 states the covariance of real operations under complex conjugation
of quantum states. That is, complex conjugation and any real operation
commute. Proposition 1 will be useful in the proof of Theorem 1.

\emph{Proposition 2.} For any real operation $\phi =\{K_{l}\}_{l}$ under the
orthonormal basis $\{|j\rangle \}_{j=1}^{d},$ there exists a real matrix $%
K^{\prime }$ in $\{|j\rangle \}_{j=1}^{d}$ such that $\phi ^{\prime
}=\{K^{\prime }\}\cup \{K_{l}\}_{l}$ becomes a real channel.

\emph{Proof.} $\phi =\{K_{l}\}_{l}$ is a real operation, then $%
I-\sum_{l}K_{l}^{\dagger }K_{l}$ is real in basis $\{|j\rangle \}_{j=1}^{d}$
and is positive semidefinite. Applying the Cholesky factorization (see for
example Corollary 7.2.9 in \cite{Horn-2013-book}), there exists a real
matrix in basis $\{|j\rangle \}_{j=1}^{d}$ such that $I-\sum_{l}K_{l}^{%
\dagger }K_{l}=(K^{\prime })^{T}K^{\prime }.$ Proposition 2 then follows.
$\hfill\blacksquare$

Because of Proposition 2, we see that in condition (M3), if we replace the real
channel by a real operation $\phi =\{K_{l}\}_{l}$ and $\sum_{l}K_{l}^{%
\dagger }K_{l}\neq I,$ then there exists a real channel $\phi ^{\prime
}=\{K^{\prime }\}\cup \{K_{l}\}_{l}$ with $K^{\prime }\neq 0$, and
\begin{eqnarray}
&&\sum_{l}\text{tr}(K_{l}\rho K_{l}^{T})M[\frac{K_{l}\rho K_{l}^{T}}{\text{tr%
}(K_{l}\rho K_{l}^{T})}] \nonumber \\
&\leq &\text{tr}[K^{\prime }\rho (K^{\prime })^{T}]M[\frac{K^{\prime }\rho
(K^{\prime })^{T}}{\text{tr}[K^{\prime }\rho (K^{\prime })^{T}]}]  \nonumber \\
&& \ \ +\sum_{l}%
\text{tr}(K_{l}\rho K_{l}^{T})M[\frac{K_{l}\rho K_{l}^{T}}{\text{tr}%
(K_{l}\rho K_{l}^{T})}]  \nonumber \\
&\leq &M(\rho ).  \nonumber
\end{eqnarray}
This implies that, it is equivalent when replacing \textquotedblleft real
channel" in (M2) by \textquotedblleft real operation".

\emph{Proposition 3.} For the given state $\rho ,$ there exists a real
orthogonal matrix $Q$ in $\{|j\rangle \}_{j=1}^{d}$ such that $Q($Re$\rho
)Q^{T}$ is diagonal.

\emph{Proof.} In Eq. (\ref{eq1-2}), since Re$\rho $ is real and symmetric, then Proposition 
3 evidently holds.$\hfill \blacksquare $

We discuss some implications of $Q\rho Q^{T}.$ With $Q$ in Proposition 3 and Eq.
(\ref{eq1-2}), we have
\begin{eqnarray}
Q\rho Q^{T}=Q(\text{Re}\rho )Q^{T}+iQ(\text{Im}\rho )Q^{T}.   \label{eq2-2}
\end{eqnarray}%
We see that $Q\rho Q^{T}$ is a density operator, and all off-diagonal
entries of $Q\rho Q^{T}$ are pure imaginary (or zero), hence $Q\rho Q^{T}$
is real if and only if $Q\rho Q^{T}$ is diagonal (incoherent in coherence
theory  \cite{PRL-Plenio-2014,RMP-Plenio-2017}). Since $Q$ and $Q^{T}=Q^{-1}$ are all real, then (M2) implies that $%
\rho $ and $Q\rho Q^{T}$ have the same imaginarity for any imaginarity measure.
For the imaginarity measure $M_{\text{r}}(\rho )$ in Eq. (\ref{eq1-5}), we have
\begin{eqnarray}
&&M_{\text{r}}(\rho )=M_{\text{r}}(Q\rho Q^{T})   \nonumber \\
&=&S[\text{Re}(Q\rho
Q^{T})]-S(Q\rho Q^{T})=C_{\text{r}}(Q\rho Q^{T}),   \label{eq2-3}
\end{eqnarray}%
where $C_{\text{r}}$ is the relative entropy of coherence \cite{PRL-Plenio-2014}. Eq. (\ref{eq2-3})
then established the equivalence between $M_{\text{r}}(\rho )$ and $C_{\text{%
r}}(Q\rho Q^{T}).$ We also see that, since $Q(\text{Re}\rho )Q^{T}$ is the diagonal part of $Q\rho Q^{T}$, then $Q(\text{Re}\rho )Q^{T}$ is a density operator, and then $\text{Re}\rho$ is still a density operator.

\section{An imaginarity measure based on Tsallis relative entropy}

Tsallis relative entropy \cite{JMP-1998-Tsallis,JMP-2004-Kuriyama,PRA-2003-Abe,PLA-2003-Abe} has many applications in quantum information science. For example, Tsallis relative entropy was used to construct coherence measures \cite{PRA-2016-Rastegin,PRA-2017-Yu,SR-2018-Yu,PRA-2020-Xu}. We now propose an imaginarity measure based on Tsallis relative entropy in Theorem 1 below.

\emph{Theorem 1. }$M_{T,\mu }(\rho )$ defined as
\begin{eqnarray}
M_{T,\mu }(\rho )=1-\text{tr}[\rho ^{\mu }(\rho ^{\ast })^{1-\mu }]  \label{eq3-1}
\end{eqnarray}
is an imaginarity measure, where $\mu \in (0,1).$

$Proof.$ We prove that $M_{T,\mu }(\rho )$ satisfies (M5), (M1) and
(M2). We first prove that $M_{T,\mu }(\rho )$ satisfies (M5).
For the direct sum state
\begin{eqnarray}
\rho =p\rho _{1}\oplus (1-p)\rho _{2} \label{eq3-2}
\end{eqnarray}
with $p\in (0,1)$ and $\{\rho _{1},\rho _{2}\}$ quantum states, we have
\begin{eqnarray}
\rho ^{\mu } &=&p^{\mu }\rho _{1}^{\mu }\oplus (1-p)^{\mu }\rho _{2}^{\mu }, \nonumber
\\
(\rho ^{\ast })^{1-\mu } &=&p^{1-\mu }(\rho _{1}^{\ast })^{1-\mu }\oplus
(1-p)^{1-\mu }(\rho _{2}^{\ast })^{1-\mu }.  \nonumber
\end{eqnarray}
It follows that
\begin{eqnarray}
&&\text{tr}[\rho ^{\mu }(\rho ^{\ast })^{1-\mu }]  \nonumber  \\
&=&\text{tr}[p\rho _{1}^{\mu }(\rho _{1}^{\ast })^{1-\mu }\oplus (1-p)\rho
_{2}^{\mu }(\rho _{2}^{\ast })^{1-\mu }]  \nonumber  \\
&=&p\text{tr}[\rho _{1}^{\mu }(\rho _{1}^{\ast })^{1-\mu }]+(1-p)\text{tr}%
[\rho _{2}^{\mu }(\rho _{2}^{\ast })^{1-\mu }],  \nonumber  \\
&&1-\text{tr}[\rho ^{\mu }(\rho ^{\ast })^{1-\mu }]  \nonumber \\
&=&p\{1-\text{tr}[\rho _{1}^{\mu }(\rho _{1}^{\ast })^{1-\mu }]\}+(1-p)\{1-%
\text{tr}[\rho _{2}^{\mu }(\rho _{2}^{\ast })^{1-\mu }]\}.  \nonumber
\end{eqnarray}
We then proved that $M_{T,\mu }(\rho )$ satisfies (M5).

To prove that $M_{T,\mu }(\rho )$ satisfies (M1) and (M2),
we consider the Tsallis relative entropy. For $\mu \in (0,1),$ the Tsallis
relative entropy of states $\rho $ and $\sigma $ is defined as
\cite{JMP-1998-Tsallis,JMP-2004-Kuriyama,PRA-2003-Abe,PLA-2003-Abe}
\begin{eqnarray}
D_{\mu }(\rho ||\sigma )=\frac{1-\text{tr}(\rho ^{\mu }\sigma ^{1-\mu })}{%
1-\mu }.  \label{eq3-3}
\end{eqnarray}
Note that Tsallis relative entropy $D_{\mu }(\rho ||\sigma )$ can be defined
for $\mu >1,$ but $D_{\mu }(\rho ||\sigma )$ may be infinite for some states
$\rho $ and $\sigma $ when $\mu >1.$ In this work, we only consider the case
$\mu \in (0,1).$ When $\mu \in (0,1),$ it is shown that
\cite{PRA-2016-Rastegin}
\begin{eqnarray}
D_{\mu }(\rho ||\sigma )\geq 0,  \label{eq3-4}
\end{eqnarray}
with equality if and only if $\rho =\sigma.$ Let $\sigma =\rho ^{\ast
}, $ it follows that $M_{T,\mu }(\rho )$ satisfies (M1).

$D_{\mu }(\rho ||\sigma )$ is nonincreasing under a quantum channel $\phi $
\cite{PRA-2016-Rastegin,RMP-2011-Petz} when $\mu \in (0,1)$, that is,
\begin{eqnarray}
D_{\mu }[\phi (\rho )||\phi (\sigma )]\leq D_{\mu }(\rho ||\sigma )  \label{eq3-5}
\end{eqnarray}
for any quantum channel $\phi $ and any states $\rho $ and $\sigma .$ Let $%
\sigma =\rho ^{\ast }$ and $\phi $ be a real channel, applying Proposition 1, we
get
\begin{eqnarray}
&&D_{\mu }[\phi (\rho )||\phi (\rho ^{\ast })]  \nonumber \\
&=&\frac{1-\text{tr}\{[\phi (\rho )]^{\mu }[\phi (\rho ^{\ast })]^{1-\mu }\}%
}{1-\mu }  \nonumber \\
&=&\frac{1-\text{tr}\{[\phi (\rho )]^{\mu }[(\phi (\rho ))^{\ast }]^{1-\mu
}\}}{1-\mu }=\frac{M_{T,\mu }[\phi (\rho )]}{1-\mu }  \nonumber \\
&\leq &\frac{1-\text{tr}[\rho ^{\mu }(\rho ^{\ast })^{1-\mu }]}{1-\mu }=%
\frac{M_{T,\mu }(\rho )}{1-\mu }. \nonumber
\end{eqnarray}
It follows that $M_{T,\mu }(\rho )$ satisfies (M2), and we
completed the proof of Theorem 1. $\hfill\blacksquare$

When $\mu =\frac{1}{2},$ $M_{T,\mu }(\rho )$ reads
\begin{eqnarray}
M_{T,\frac{1}{2}}(\rho )=1-\text{tr}(\sqrt{\rho }\sqrt{\rho ^{\ast }}).  \label{eq3-6}
\end{eqnarray}

In the definition of $M_{T,\mu }(\rho )$ in Eq. (\ref{eq3-1}), since tr$[\rho ^{\mu
}(\rho ^{\ast })^{1-\mu }]$ is real, then it is obvious that
\begin{eqnarray}
M_{T,\mu }(\rho )=M_{T,\mu }(\rho ^{\ast })=M_{T,1-\mu }(\rho ).  \label{eq3-7}
\end{eqnarray}
Further, when $\rho =|\psi \rangle \langle \psi |$ is a pure state,
\begin{eqnarray}
M_{T,\mu }(|\psi \rangle \langle \psi |)=1-|\langle \psi |\psi ^{\ast
}\rangle |^{2},  \label{eq3-8}
\end{eqnarray}
this compares with $M_{F}(|\psi \rangle \langle \psi |)$ in Eq.
(\ref{eq1-3}) that
\begin{eqnarray}
M_{F}(|\psi \rangle \langle \psi |)=1-|\langle \psi |\psi ^{\ast }\rangle |.  \label{eq3-9}
\end{eqnarray}

\emph{Example 1.} $M_{T,\mu }(\rho )$ for qubit states. Any qubit state $%
\rho $ can be expressed in Bloch representation as
\begin{eqnarray}
\rho (\overrightarrow{r})=\frac{1}{2}(I+\overrightarrow{r}\cdot
\overrightarrow{\sigma })=\frac{1}{2}\left(
\begin{array}{cc}
1+z & x-iy \\
x+iy & 1-z%
\end{array}%
\right) ,  \label{eq3-10}
\end{eqnarray}
where $\overrightarrow{r}\cdot \overrightarrow{\sigma }=x\sigma _{x}+y\sigma
_{y}+z\sigma _{z},$ $\overrightarrow{r}=(x,y,z)$ is a real vector with $r=|%
\overrightarrow{r}|=\sqrt{x^{2}+y^{2}+z^{2}}\leq 1,$ $\overrightarrow{\sigma
}=(\sigma _{x},\sigma _{y},\sigma _{z})$ are Pauli matrices $\sigma
_{x}=\left(
\begin{array}{cc}
0 & 1 \\
1 & 0%
\end{array}%
\right) ,$ $\sigma _{y}=\left(
\begin{array}{cc}
0 & -i \\
i & 0%
\end{array}%
\right) ,$ $\sigma _{z}=\left(
\begin{array}{cc}
1 & 0 \\
0 & -1%
\end{array}%
\right) .$ For $\rho (\overrightarrow{r})$ expressed in Eq. (\ref{eq3-10}), $M_{T,\mu
}[\rho (\overrightarrow{r})]$ reads
\begin{eqnarray}
&&M_{T,\mu }[\rho (\overrightarrow{r})]     \nonumber  \\
&=&1-\frac{1}{2r^{2}}\{(1-r)[(r-\frac{%
y^{2}}{r-z})^{2}+\frac{x^{2}y^{2}}{(r-z)^{2}}]     \nonumber \\
&&+(1+r)[(r-\frac{y^{2}}{r+z})^{2}+\frac{x^{2}y^{2}}{(r+z)^{2}}]     \nonumber \\
&&+y^{2}[(1-r)^{\mu }(1+r)^{1-\mu }+(1-r)^{1-\mu }(1+r)^{\mu }]\}.  \ \ \ \    \label{eq3-11}
\end{eqnarray}

We provide a proof for Eq. (\ref{eq3-11}) in Appendix A.

In particular, when $\mu =\frac{1}{2},$
\begin{eqnarray}
&&M_{T,\frac{1}{2}}[\rho (\overrightarrow{r})]   \nonumber  \\
&=&1-\frac{1}{2r^{2}}\{(1-r)[(r-%
\frac{y^{2}}{r-z})^{2}+\frac{x^{2}y^{2}}{(r-z)^{2}}]    \ \ \ \ \ \   \nonumber  \\
&&+(1+r)[(r-\frac{y^{2}}{r+z})^{2}+\frac{x^{2}y^{2}}{(r+z)^{2}}]  \nonumber  \\
&&+2y^{2}\sqrt{1-r^{2}}]\}.    \label{eq3-12}
\end{eqnarray}
Let $x=z=0\leq y\leq 1,$ then Eqs. (\ref{eq3-11},\ref{eq3-12}) yield
\begin{eqnarray}
&&M_{T,\mu }(\rho )   \nonumber  \\
&=&1-\frac{1}{2}[(1+y)^{\mu }(1-y)^{1-\mu }
+(1-y)^{\mu}(1+y)^{1-\mu }].    \ \ \ \ \   \label{eq3-13}\\
&&M_{T,\frac{1}{2}}(\rho )=1-\sqrt{1-y^{2}}.   \label{eq3-14}
\end{eqnarray}
We depict Eq. (\ref{eq3-13}) in Fig. 1.

\begin{figure}
\includegraphics[width=8cm]{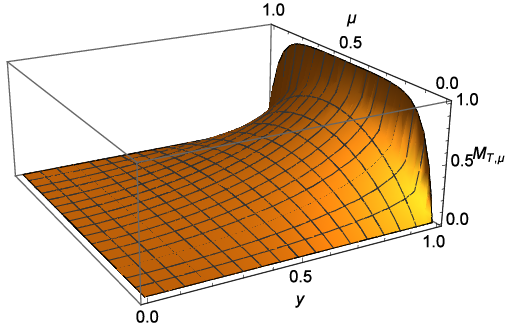}
\caption{$M_{T,\mu }(\rho )$ versus $(y,\mu)$ in Eq. (\ref{eq3-13}).}
\end{figure}

\section{Linking $M_{T,\mu }(\rho )$ to other quantities}

For two density operators $\rho $ and $\sigma ,$ the quantum affinity $%
A(\rho ,\sigma )$ is defined as cite{}
\begin{eqnarray}
A(\rho ,\sigma )=\text{tr}(\sqrt{\rho }\sqrt{\sigma });
\end{eqnarray}
the quantum Hellinger distance $D_{\text{H}}(\rho ,\sigma )$ is defined as
cite{}
\begin{eqnarray}
D_{\text{H}}(\rho ,\sigma )=\text{tr}(\sqrt{\rho }-\sqrt{\sigma }%
)^{2}=2[1-A(\rho ,\sigma )];
\end{eqnarray}
the quantum Bhattacharyya bound $P_{\text{B}}(\rho,\sigma)$ is defined
as cite{}
\begin{eqnarray}
P_{\text{B}}(\rho,\sigma)=\frac{1}{2}\text{tr}(\sqrt{\rho }\sqrt{%
\sigma })=\frac{1}{2}A(\rho ,\sigma );
\end{eqnarray}
and the quantum Chernoff bound $P_{\text{C}}(\rho,\sigma)$ is defined
as cite{}
\begin{eqnarray}
P_{\text{C}}(\rho,\sigma)=\frac{1}{2}\inf_{0\leq \mu \leq 1}C_{\mu
}, \text{with} \  C_{\mu }=\text{tr}(\rho ^{\mu }\sigma ^{1-\mu }).
\end{eqnarray}
Together with the definitions of $M_{T,\mu }(\rho )$ in Eq. (\ref{eq3-1}) and $M_{T,%
\frac{1}{2}}(\rho )$ in Eq. (\ref{eq3-6}), we can easily find the relations
\begin{eqnarray}
M_{T,\frac{1}{2}}(\rho ) &=&1-A(\rho ,\rho ^{\ast })=\frac{1}{2}D_{\text{H}%
}(\rho ,\rho ^{\ast }) \nonumber \\
&=&1-2P_{\text{B}}(\rho,\rho ^{\ast }); \\
P_{\text{C}}(\rho,\rho ^{\ast }) &=&\frac{1}{2}\inf_{0\leq \mu \leq
1}M_{T,\mu }(\rho ).
\end{eqnarray}
$M_{T,\mu }(\rho )$ and $M_{T,\frac{1}{2}}(\rho )$ then obtain the
respective interpretations via the interpretations of $A(\rho ,\rho
^{\ast }),$ $D_{\text{H}}(\rho ,\rho ^{\ast }),$ $P_{\text{B}}(\rho,\rho
^{\ast })$ and $P_{\text{C}}(\rho,\rho ^{\ast }).$

\section{$M_{T,\mu }(\rho )$ for Gaussian states}

Gaussian states in bosonic systems are widely used in quantum optics and quantum information theory \cite{RMP-2005-Braunstein,PR-2007-Wang,arXiv-2005-Ferraro,EPJ-2012-Olivares,
RMP-2012-Weedbrook,OSID-2014-Adesso,book-2017-Serafini}. Imaginarity theory is dependent on an orthonormal basis. Fock basis is the orthonormal basis of the Hilbert space for Gaussian states, then it is natural to choose Fock basis to establish the imaginarity theory of Gaussian states. However, it is convenient to express Gaussian states by the means and covariance matrices, and difficult to express general Gaussian states in Fock basis (some explorations about expressing general Gaussian states in Fock basis see for examples \cite{PRA-Xu-2016,PRA-2019-Quesada,JPA-2020-Huh,arxiv-2022-Yao,arXiv-2023-Xu}). Therefore, there arises a problem that whether we can express an imaginarity measure of Gaussian states by the means and covariance matrices. In this section, we give the expression of $M_{T,\mu }(\rho )$ for Gaussian
states in terms of the means and covariance matrices of Gaussian states.

We first review some basics of Gaussian states and introduce the
notation we will use. Suppose the $N$-mode Fock basis is $\{|j\rangle
\}_{j}^{\otimes N}$ which is the $N$-fold tensor product of one-mode Fock
basis $\{|j\rangle \}_{j=0}^{\infty }.$ Suppose each $\{|j\rangle
\}_{j=0}^{\infty }$ spans the complex Hilbert space $\overline{H},$ then $%
\{|j\rangle \}_{j}^{\otimes N}$ spans the complex Hilbert space $\overline{H}%
^{\otimes N}.$ Fock basis $\{|j\rangle
\}_{j}^{\otimes N}$ is the orthonormal basis of $\overline{H}%
^{\otimes N},$ thus we fix Fock basis as reference basis in imaginarity theory.

On each $\overline{H}_{l},$ the annihilation operator $\widehat{a}_{l}$ and
creation operator $\widehat{a}_{l}^{\dagger }$ are defined as
\begin{eqnarray}
\widehat{a}_{l}|0\rangle &=&0,\ \ \widehat{a}_{l}|j\rangle =\sqrt{j}%
|j-1\rangle \text{ for }j\geq 1;  \label{eq4-1} \\
\widehat{a}_{l}^{\dagger }|j\rangle &=&\sqrt{j+1}|j+1\rangle \text{ for }%
j\geq 0.  \label{eq4-2}
\end{eqnarray}%
From $\{\widehat{a}_{l},\widehat{a}_{l}^{\dagger }\}_{l=1}^{N}$ we can
define the quadrature field operators $\{\widehat{q}_{l},\widehat{p}%
_{l}^{\dagger }\}_{l=1}^{N}$ as
\begin{eqnarray}
\widehat{q}_{l}=\widehat{a}_{l}+\widehat{a}_{l}^{\dagger },\ \ \widehat{p}%
_{l}=-i(\widehat{a}_{l}-\widehat{a}_{l}^{\dagger }).  \label{eq4-3}
\end{eqnarray}%
We organize $\{\widehat{q}_{l},\widehat{p}_{l}^{\dagger }\}_{l=1}^{N}$ as a
vector
\begin{eqnarray}
\widehat{X} &=&(\widehat{q}_{1},\widehat{p}_{1},\widehat{q}_{2},\widehat{p}%
_{2},...,\widehat{q}_{N},\widehat{p}_{N})^{T}  \notag \\
&=&(\widehat{X}_{1},\widehat{X}_{2},\widehat{X}_{3},\widehat{X}_{4},...,%
\widehat{X}_{2N-1},\widehat{X}_{2N})^{T}.  \label{eq4-4}
\end{eqnarray}

For state $\rho $ in $\overline{H}^{\otimes N},$ the mean of $\rho $ is
defined as
\begin{eqnarray}
\overline{X}=\text{tr}(\rho \widehat{X})=(\overline{X}_{1},\overline{X}%
_{2},...,\overline{X}_{2N})^{T};  \label{eq4-5}
\end{eqnarray}%
the covariance matrix $V$ is defined by its entries
\begin{eqnarray}
V_{lm}=\frac{1}{2}\text{tr}(\rho \{\Delta \widehat{X}_{l},\Delta \widehat{X}%
_{m}\}),  \label{eq4-6}
\end{eqnarray}%
where $\Delta \widehat{X}_{l}=\widehat{X}_{l}-\overline{X}_{l},$ $\{\Delta
\widehat{X}_{l},\Delta \widehat{X}_{m}\}=\Delta \widehat{X}_{l}\Delta
\widehat{X}_{m}+\Delta \widehat{X}_{m}\Delta \widehat{X}_{l}.$ The
covariance matrix $V$ is a $2N\times 2N$ real symmetric matrix and
fulfills the uncertainty principle \cite{PRA-1994-Simon}
\begin{eqnarray}
V+i\Omega \succeq 0,  \label{eq4-7}
\end{eqnarray}
where
\begin{eqnarray}
\Omega =\oplus _{l=1}^{N}\left(
\begin{array}{cc}
0 & 1 \\
-1 & 0%
\end{array}%
\right) .   \label{eq4-8}
\end{eqnarray}

For a quantum state $\rho $ in $\overline{H}^{\otimes N},$ its
characteristic function is defined as
\begin{eqnarray}
\chi (\rho ,\xi )=\text{tr}[\rho D(\xi )],  \label{eq4-9}
\end{eqnarray}%
where $D(\xi )$ is the displacement operator
\begin{eqnarray}
D(\xi )=\exp (i\widehat{X}^{T}\Omega \xi ),  \label{eq4-10}
\end{eqnarray}
$\xi =(\xi _{1},\xi _{2},...,\xi _{2N})^{T}$ is a real vector.

A quantum state $\rho $ in $\overline{H}^{\otimes N}$ is said to be an $N$%
-mode Gaussian state if its characteristic function possesses the Gaussian
form
\begin{eqnarray}
\chi (\rho ,\xi )=\exp [-\frac{1}{2}\xi ^{T}(\Omega V\Omega ^{T})\xi
-i(\Omega \overline{X})^{T}\xi ]  \label{eq4-11}
\end{eqnarray}%
with $\overline{X}$ and $V$ the mean and covariance matrix of $\rho .$ $%
\overline{X}$ and $V$ with Eq. (\ref{eq4-7}) fully determine the
Gaussian state $\rho $ \cite{PRA-1994-Simon}, hence we write $\rho $ as $%
\rho (\overline{X},V).$

With these preparations, we state Proposition 4 and Theorem 2 as follows.

\emph{Proposition 4.} Suppose $\rho \left( \overline{X},V\right) $ is an $N$%
-mode Gaussian state and $\mu >0.$ Then $\frac{\rho ^{\mu }}{\text{tr}\rho
^{\mu }}$ is still a Gaussian state. The mean of $\frac{\rho ^{\mu }}{\text{%
tr}\rho ^{\mu }}$ is still $\overline{X}.$ We denote the covariance matrix
of $\frac{\rho ^{\mu }}{\text{tr}\rho ^{\mu }}$ by $V^{(\mu )},$ $V^{(\mu )}$
can be obtained as follows.
\begin{eqnarray}
V &=&S(\oplus _{l=1}^{N}\nu _{l}I_{2})S^{T},\nu _{l}\geq 1,   \label{eq4-12}\\
\nu _{l}^{(\mu )} &=&\frac{2}{1-(\frac{\nu _{l}-1}{\nu _{l}+1})^{\mu }}%
-1,1\leq l\leq N,   \label{eq4-13}\\
V^{(\mu )} &=&S(\oplus _{l=1}^{N}\nu _{l}^{(\mu )}I_{2})S^{T},  \label{eq4-14}
\end{eqnarray}
where $I_{2}$ is the identity matrix of size $2,$ $S$ is a $2N\times 2N$
real matrix, $S$ is also symplectic i.e. $S$ satisfies $S\Omega S^{T}=\Omega.$

\emph{Proof.} Proposition 5 is equivalent to Lemma 1 in Ref. \cite{PRA-2008-Lloyd}, we provide this proof for completeness. Any $N$-mode Gaussian state $\rho \left( \overline{X},V\right)
$ can be written in the form of thermal decomposition as (see for example
chapter 3 in Ref. \cite{book-2017-Serafini})
\begin{eqnarray}
\rho \left( \overline{X},V\right) =D(\frac{\overline{X}}{2})U_{S}[\otimes
_{l=1}^{N}\rho _{\text{th}}(\frac{\nu _{l}-1}{2})]U_{S}^{\dagger }D(-\frac{%
\overline{X}}{2}),   \label{eq4-15}
\end{eqnarray}
where $U_{S}$ is a unitary operation associated with the real $2N\times 2N$
real symplectic matrix $S,$
\begin{eqnarray}
&&\rho _{\text{th}}(\frac{\nu _{l}-1}{2})=(1-e^{-\eta
_{l}})\sum_{j=0}^{\infty }e^{-j\eta _{l}}|j\rangle \langle j|,   \label{eq4-16}\\
&&\nu _{l}=\frac{1+e^{-\eta _{l}}}{1-e^{-\eta _{l}}}\geq 1\text{ or }\eta
_{l}=\ln \frac{\nu _{l}+1}{\nu _{l}-1}\geq 0,  \label{eq4-17}
\end{eqnarray}%
$\rho _{\text{th}}(\frac{\nu _{l}-1}{2})$ is a thermal state with mean
number $\frac{\nu _{l}-1}{2}=$tr$[\widehat{a}_{l}^{\dagger }\widehat{a}%
_{l}\rho _{\text{th}}(\frac{\nu _{j}-1}{2})],$ $S$ and $\nu _{l}$ are
obtained by Eq. (\ref{eq4-12}). From Eq. (\ref{eq4-15}), we get
\begin{eqnarray}
\frac{\rho ^{\mu }}{\text{tr}\rho ^{\mu }} &=&D(\frac{\overline{X}}{2}%
)U_{S}\{\otimes _{l=1}^{N}[(1-e^{-\mu \eta _{l}})\sum_{j=0}^{\infty
}e^{-j\mu \eta _{l}}|j\rangle \langle j|]\}  \nonumber \\
&&\cdot U_{S}^{\dagger }D(-\frac{%
\overline{X}}{2})  \nonumber \\
&=&D(\frac{\overline{X}}{2})U_{S}[\otimes _{l=1}^{N}\rho _{\text{th}}(\frac{%
\nu _{l}^{(\mu )}-1}{2})]U_{S}^{\dagger }D(-\frac{\overline{X}}{2}), \label{eq4-18}
\end{eqnarray}
where
\begin{eqnarray}
\nu _{l}^{(\mu )}=\frac{1+e^{-\mu \eta _{l}}}{1-e^{-\mu \eta _{l}}}. \label{eq4-19}
\end{eqnarray}
Then Proposition 4 follows.
$\hfill\blacksquare$

Due to Proposition 4, if $\rho \left( \overline{X},V\right) $ is an $N$-mode
Gaussian state, then $\frac{\rho ^{\mu }}{\text{tr}\rho ^{\mu }}$ is a
Gaussian state with $\left( \overline{X},V^{(\mu )}\right) $ its mean and
covariance matrix, we thus write $\frac{\rho ^{\mu }}{\text{tr}\rho ^{\mu }}$
as $\frac{\rho ^{\mu }}{\text{tr}\rho ^{\mu }}\left( \overline{X},V^{(\mu
)}\right) .$ It is shown that \cite{arXiv-2023-Xu} if $%
\rho \left( \overline{X},V\right) $ is an $N$-mode Gaussian state, then its
complex conjugate, $\rho ^{\ast },$ is also a Gaussian state with mean $%
\overline{X}^{\prime }$ and covariance matrix $V^{\prime }$ as
\begin{eqnarray}
\overline{X}^{\prime } &=&O\overline{X},\text{\ \ }V^{\prime }=OVO,  \label{eq4-20} \\
O &=&\oplus _{l=1}^{N}\left(
\begin{array}{cc}
1 & 0 \\
0 & -1%
\end{array}%
\right) .   \label{eq4-21}
\end{eqnarray}
Thus we write $\rho ^{\ast }$ as $\rho ^{\ast }(\overline{X}^{\prime
},V^{\prime }).$ $\rho \left( \overline{X},V\right) $ is a real Gaussian state if and only if $\{\overline{X}_{l}=0\}_{l=1}^{N}$ and $\{V_{2l-1,2m}=0\}_{l,m=1}^{N}$  \cite{arXiv-2023-Xu}. From Eqs. (\ref{eq4-12},\ref{eq4-20},\ref{eq4-21}) we have
\begin{eqnarray}
V^{\prime } &=&OS(\oplus _{l=1}^{N}\nu _{l}I_{2})S^{T}O    \nonumber \\
&=&(OSO)(\oplus _{l=1}^{N}\nu _{l}I_{2})(OSO)^{T},  \label{eq4-22}
\end{eqnarray}
where we have used the fact $\oplus _{l=1}^{N}\nu _{l}I_{2}=O(\oplus
_{l=1}^{N}\nu _{l}I_{2})O.$ Because of the facts $O^{2}=I_{2N},$ $O\Omega =-\Omega O,$ and $S\Omega S^{T}=\Omega,$ here $I_{2N}$ is the identity matrix of size $2N,$ we have
\begin{eqnarray}
(OSO)\Omega (OSO)^{T}=\Omega.   \label{eq4-23}
\end{eqnarray}
That is, $OSO$ is a real and symplectic matrix. Employing Proposition 4, we see
that $\frac{(\rho ^{\ast })^{(1-\mu )}}{\text{tr}(\rho ^{\ast })^{(1-\mu )}}$
is also a Gaussian state with mean $\overline{X}^{\prime }$ and covariance
matrix $V^{\prime (1-\mu )}.$ $V^{\prime (1-\mu )}$ can be obtained as
follows,
\begin{eqnarray}
\nu _{l}^{(1-\mu )} &=&\frac{2}{1-(\frac{\nu _{l}-1}{\nu _{l}+1})^{1-\mu }}%
-1,1\leq l\leq N,            \label{eq4-24}\\
V^{\prime (1-\mu )} &=&OS(\oplus _{l=1}^{N}\nu _{l}^{(1-\mu
)}I_{2})S^{T}O=OV^{(1-\mu )}O.   \ \  \label{eq4-25}
\end{eqnarray}
We thus write $\frac{(\rho ^{\ast })^{(1-\mu )}}{\text{tr}(\rho ^{\ast
})^{(1-\mu )}}$ as $\frac{(\rho ^{\ast })^{(1-\mu )}}{\text{tr}(\rho ^{\ast
})^{(1-\mu )}}\left( \overline{X}^{\prime },V^{\prime (1-\mu )}\right) .$
\begin{widetext}
Theorem 2 below gives the expression of $M_{T,\mu }(\rho )$ for any $N$-mode
Gaussian state in terms of mean and covariance matrix.

\emph{Theorem 2.} For $N$-mode Gaussian state $\rho \left( \overline{X}%
,V\right) $ and $\mu \in (0,1),$ $M_{T,\mu }(\rho )$ defined in Eq. (\ref{eq3-1}) has
the expression
\begin{eqnarray}
M_{T,\mu }(\rho )=1-\frac{2^{N}\Pi _{l=1}^{N}\frac{1-e^{-\eta _{l}}}{%
(1-e^{-\mu \eta _{l}})(1-e^{-(1-\mu )\eta _{l}})}}{\sqrt{\det [V^{(\mu
)}+V^{\prime (1-\mu )}]}}\exp \{-\frac{1}{2}(\overline{X}-\overline{X}%
^{\prime })^{T}[V^{(\mu )}+V^{\prime (1-\mu )}]^{-1}(\overline{X}-\overline{X%
}^{\prime })\},                 \label{eq4-26}
\end{eqnarray}
where $\eta _{l},$ $V^{(\mu )},$ $V^{\prime (1-\mu
)},$  $\overline{X}^{\prime },$ are defined in Eqs. (\ref{eq4-17},\ref{eq4-14},\ref{eq4-25},\ref{eq4-20}).

\emph{Proof.} This proof is inspired by the proof of Theorem 2 in Ref. \cite{PRA-2008-Lloyd}. The overlap tr$(\rho \sigma )$ of two $N$-mode Gaussian states
$\rho \left( \overline{X},V\right) $ and $\sigma \left( \overline{Y}%
,W\right) $ has the expression (see for example chapter 3 in Ref.
\cite{book-2017-Serafini})
\begin{eqnarray}
\text{tr}(\rho \sigma )=\frac{2^{N}}{\sqrt{\det (V+W)}}\exp [-\frac{1}{2}(%
\overline{X}-\overline{Y})^{T}(V+W)^{-1}(\overline{X}-\overline{Y})].             \label{eq4-27}
\end{eqnarray}
Taking the two $N$-mode Gaussian states $\frac{\rho ^{\mu }}{\text{tr}\rho
^{\mu }}\left( \overline{X},V^{(\mu )}\right) $ and $\frac{(\rho ^{\ast
})^{(1-\mu )}}{\text{tr}(\rho ^{\ast })^{(1-\mu )}}\left( \overline{X}%
^{\prime },V^{\prime (1-\mu )}\right) $ into Eqs. (\ref{eq4-27},\ref{eq3-1}), we get
\begin{eqnarray}
\text{tr}[\rho ^{\mu }(\rho ^{\ast })^{1-\mu }]
&=&[\text{tr}\rho ^{\mu }][\text{tr}(\rho ^{\ast })^{1-\mu }]\text{tr}[\frac{%
\rho ^{\mu }}{\text{tr}\rho ^{\mu }}\frac{(\rho ^{\ast })^{1-\mu }}{\text{tr}%
(\rho ^{\ast })^{1-\mu }}],        \nonumber  \\
\text{tr}\rho ^{\mu } &=&\Pi _{l=1}^{N}\frac{(1-e^{-\eta _{l}})^{\mu }}{%
1-e^{-\mu \eta _{l}}},        \nonumber  \\
\text{tr}(\rho ^{\ast })^{1-\mu } &=&\text{tr}\rho ^{1-\mu }=\Pi _{l=1}^{N}%
\frac{(1-e^{-\eta _{l}})^{1-\mu }}{1-e^{-(1-\mu )\eta _{l}}},        \nonumber  \\
\lbrack \text{tr}\rho ^{\mu }][\text{tr}(\rho ^{\ast })^{1-\mu }] &=&\Pi
_{l=1}^{N}\frac{1-e^{-\eta _{l}}}{(1-e^{-\mu \eta _{l}})(1-e^{-(1-\mu )\eta
_{l}})},        \nonumber  \\
\text{tr}[\frac{\rho ^{\mu }}{\text{tr}\rho ^{\mu }}\frac{(\rho ^{\ast
})^{1-\mu }}{\text{tr}(\rho ^{\ast })^{1-\mu }}] &=&\frac{2^{N}}{\sqrt{\det
[V^{(\mu )}+V^{\prime (1-\mu )}]}}\exp \{-\frac{1}{2}(\overline{X}-\overline{%
X}^{\prime })^{T}[V^{(\mu )}+V^{\prime (1-\mu )}]^{-1}(\overline{X}-%
\overline{X}^{\prime })\}.   \label{eq4-28}
\end{eqnarray}
We then proved Theorem 2.
$\hfill\blacksquare$

In particular, when $\mu =\frac{1}{2},$
\begin{eqnarray}
M_{T,\frac{1}{2}}(\rho )=1-\frac{2^{N}\Pi _{l=1}^{N}\frac{1-e^{-\eta _{l}}}{%
(1-e^{-\frac{1}{2}\eta _{l}})^{2}}}{\sqrt{\det [V^{(\frac{1}{2})}+OV^{(\frac{%
1}{2})}O]}}\exp \{-\frac{1}{2}(\overline{X}-\overline{X}^{\prime
})^{T}[V^{(\frac{1}{2})}+OV^{(\frac{1}{2})}O]^{-1}(\overline{X}-\overline{X}%
^{\prime })\}.       \label{eq4-29}
\end{eqnarray}

\emph{Example 2.} $M_{T,\frac{1}{2}}(\rho )$ for one-mode Gaussian states. For
one-mode case, we write $\widehat{a}_{1}=\widehat{a},$ $\widehat{a}%
_{1}^{\dagger }=\widehat{a}^{\dagger },$ $\nu _{1}=\nu ,$ $\eta _{1}=\eta .$
Any one-mode Gaussian state $\rho \left( \overline{X},V\right) $ has the
thermal decomposition \cite{JMO-1995-Adam,EPJ-2012-Olivares}
\begin{eqnarray}
\rho \left( \overline{X},V\right) =D(\alpha )S(\zeta )\rho _{\text{th}}(%
\frac{\nu -1}{2})S(-\zeta )D(-\alpha ),    \label{eq4-30}
\end{eqnarray}%
where $\alpha ,$ $\zeta $ are complex numbers, $\zeta =|\zeta |e^{i\theta }$
is the polar form, $D(\alpha )=\exp (\alpha \widehat{a}^{\dagger }-\alpha
^{\ast }\widehat{a})$ is the one-mode displacement operator, $S(\zeta )=\exp
[\frac{1}{2}(\zeta ^{\ast }\widehat{a}^{2}-\zeta \widehat{a}^{\dagger 2})]$
is the one-mode squeezing operator, the mean $\overline{X}=2(\text{Re}\alpha ,\text{Im}\alpha ),$ 
and the covariance matrix 
\begin{eqnarray}
V=\nu \left(
\begin{array}{cc}
\cosh (2|\zeta |)+\cos \theta \sinh (2|\zeta |) & \sin \theta \sinh (2|\zeta
|) \\
\sin \theta \sinh (2|\zeta |) & \cosh (2|\zeta |)-\cos \theta \sinh (2|\zeta
|)%
\end{array}%
\right) .      \label{eq4-31}
\end{eqnarray}
For one-mode Gaussian state $\rho \left( \overline{X},V\right) $ expressed
in Eq. (\ref{eq4-30}), Eq. (\ref{eq4-29}) yields
\begin{eqnarray}
M_{T,\frac{1}{2}}(\rho )=1-\frac{\exp \{-\frac{\overline{X}_{2}^{2}}{(\nu +%
\sqrt{\nu ^{2}-1})[\cosh (2|\zeta |)-\cos \theta \sinh (2|\zeta |)]}\}}{\sqrt{%
1+\sin ^{2}\theta \sinh ^{2}(2|\zeta |)}}. \label{eq4-32}
\end{eqnarray}
We provide a proof for Eq. (\ref{eq4-32}) in Appendix B.
\end{widetext}

We discuss some properties of Eq. (\ref{eq4-31}) as follows.

(i). $M_{T,\frac{1}{2}}(\rho )=0$ when $\overline{X}_{2}=0=\sin \theta \sinh
(2|\zeta |)=0.$ This coincides with the fact that $\rho \left( \overline{X}%
,V\right) $ is real when $\overline{X}_{2}=0=\sin \theta \sinh (2|\zeta
|)=0. $

(ii). Let $\zeta $ and $\nu $ keep constant. Then $M_{T,\frac{1}{2}}(\rho )$
increases when $|\overline{X}_{2}|$ increases, $M_{T,\frac{1}{2}}(\rho )$
tends to $1$ when $|\overline{X}_{2}|$ tends to $+\infty .$

(iii). Let $\zeta $ and $\overline{X}_{2}$ keep constant, $\overline{X}%
_{2}\neq 0.$ Then $M_{T,\frac{1}{2}}(\rho )$ decreases when $\nu $ increases.

\section{Summary}

We proposed an imaginarity measure $M_{T,\mu }$ based on the Tsallis
relative entropy. $M_{T,\mu }$ has explicit expression, and also, $M_{T,\mu
} $ is computable for general Gaussian states in terms of the means and
covariances of Gaussian states.

There remained many open questions for future research. First, since $%
M_{T,\mu }$ is comparatively easy to compute, thus $M_{T,\mu }$ provide an
advantageous tool to study the dynamical behaviors of the imaginarity for both finite-dimensional states and Gaussian states. Second, finding
operational interpretations to $M_{T,\mu }$ are desirable to reveal more
implications of $M_{T,\mu }.$ Third, relationships between $M_{T,\mu }$ and
other imaginarity measures, and further, relationships between $M_{T,\mu }$
and other quantum properties, such as uncertainty relation of $M_{T,\mu }$
with respect to two orthonormal bases, are worthy of investigations.

\section*{ACKNOWLEDGMENTS}

This work was supported by the Natural Science Basic Research Plan in
Shaanxi Province of China (Program No. 2022JM-012).

\section*{Appendix A: Proof of Eq. (\ref{eq3-11})}

\label{AppendixA} \setcounter{equation}{0} \renewcommand\theequation{A%
\arabic{equation}}

The unitary diagonalization of $\rho (\overrightarrow{r})$ in Eq. (\ref{eq3-10}) is
\begin{eqnarray}
\rho &=&U\overline{\rho }U^{\dagger },   \label{eqA-1}\\
\overline{\rho } &=&\left(
\begin{array}{cc}
\frac{1-r}{2} & 0 \\
0 & \frac{1+r}{2}%
\end{array}%
\right) ,   \label{eqA-2}  \\
U &=&\left(
\begin{array}{cc}
-\sqrt{\frac{r-z}{2r}} & \sqrt{\frac{r+z}{2r}} \\
\frac{x+iy}{\sqrt{2r(r-z)}} & \frac{x+iy}{\sqrt{2r(r+z)}}%
\end{array}%
\right) .    \label{eqA-3}
\end{eqnarray}
Consequently,
\begin{eqnarray}
&&\text{tr}[\rho ^{\mu }(\rho ^{\ast })^{1-\mu }]    \nonumber  \\
&=&\text{tr}[U\overline{\rho }^{\mu }U^{\dagger }U^{\ast }\overline{\rho }%
^{1-\mu }U^{T}]     \nonumber  \\
&=&\text{tr}[(U^{T}U)\overline{\rho }^{\mu }(U^{T}U)^{\ast }\overline{\rho }%
^{1-\mu }].    \label{eqA-4}
\end{eqnarray}
Eqs. (\ref{eqA-2},\ref{eqA-3}) ensure that
\begin{widetext}
\begin{eqnarray}
U^{T}U &=&\frac{1}{2r}\left(
\begin{array}{cc}
r-z+\frac{(x+iy)^{2}}{r-z} & -\sqrt{x^{2}+y^{2}}+\frac{(x+iy)^{2}}{\sqrt{%
x^{2}+y^{2}}} \\
-\sqrt{x^{2}+y^{2}}+\frac{(x+iy)^{2}}{\sqrt{x^{2}+y^{2}}} & r+z+\frac{%
(x+iy)^{2}}{r+z}%
\end{array}%
\right) ,     \label{eqA-5} \\
U^{T}U\overline{\rho }^{\mu } &=&\frac{1}{2r}\left(
\begin{array}{cc}
(\frac{1-r}{2})^{\mu }[r-z+\frac{(x+iy)^{2}}{r-z}] & (\frac{1+r}{2})^{\mu }[%
\frac{(x+iy)^{2}}{\sqrt{x^{2}+y^{2}}}-\sqrt{x^{2}+y^{2}}] \\
(\frac{1-r}{2})^{\mu }[\frac{(x+iy)^{2}}{\sqrt{x^{2}+y^{2}}}-\sqrt{%
x^{2}+y^{2}}] & (\frac{1+r}{2})^{\mu }[r+z+\frac{(x+iy)^{2}}{r+z}]%
\end{array}%
\right) ,     \label{eqA-6} \\
(U^{T}U)^{\ast }\overline{\rho }^{1-\mu } &=&\frac{1}{2r}\left(
\begin{array}{cc}
(\frac{1-r}{2})^{1-\mu }[r-z+\frac{(x-iy)^{2}}{r-z}] & (\frac{1+r}{2}%
)^{1-\mu }[\frac{(x-iy)^{2}}{\sqrt{x^{2}+y^{2}}}-\sqrt{x^{2}+y^{2}}] \\
(\frac{1-r}{2})^{1-\mu }[\frac{(x-iy)^{2}}{\sqrt{x^{2}+y^{2}}}-\sqrt{%
x^{2}+y^{2}}] & (\frac{1+r}{2})^{1-\mu }[r+z+\frac{(x-iy)^{2}}{r+z}]%
\end{array}%
\right) ,     \label{eqA-7}\\
\text{tr}[(U^{T}U)\overline{\rho }^{\mu }(U^{T}U)^{\ast }\overline{\rho }%
^{1-\mu }]
&=&\frac{1}{4r^{2}}\{\frac{1-r}{2}|r-z+\frac{(x+iy)^{2}}{r-z}|^{2}+\frac{1+r%
}{2}|r+z+\frac{(x+iy)^{2}}{r+z}|^{2}   \nonumber  \\
&&+\frac{1}{2}[(1-r)^{1-\mu }(1+r)^{\mu }+(1-r)^{\mu }(1+r)^{1-\mu }]|\frac{%
(x+iy)^{2}}{\sqrt{x^{2}+y^{2}}}-\sqrt{x^{2}+y^{2}}|^{2}\}.     \label{eqA-8}
\end{eqnarray}
In Eq. (\ref{eqA-8}),
\begin{eqnarray}
|r-z+\frac{(x+iy)^{2}}{r-z}|^{2} &=&4(r-\frac{y^{2}}{r-z})^{2}+\frac{%
4x^{2}y^{2}}{(r-z)^{2}},   \label{eqA-9}\\
|r+z+\frac{(x+iy)^{2}}{r+z}|^{2} &=&4(r-\frac{y^{2}}{r+z})^{2}+\frac{%
4x^{2}y^{2}}{(r+z)^{2}},   \label{eqA-10}\\
|\frac{(x+iy)^{2}}{\sqrt{x^{2}+y^{2}}}-\sqrt{x^{2}+y^{2}}|^{2} &=&4y^{2}.  \label{eqA-11}
\end{eqnarray}
Taking Eqs. (\ref{eqA-9},\ref{eqA-10},\ref{eqA-11}) into Eq. (\ref{eqA-8}), Eq. (\ref{eq3-11}) then follows.

\section*{Appendix B: Proof of Eq. (\ref{eq4-31})}

\label{AppendixB} \setcounter{equation}{0} \renewcommand\theequation{B%
\arabic{equation}}

From Eqs. (\ref{eq4-13},\ref{eq4-17}), we derive $\nu ^{(\frac{1}{2})}$ and $\frac{1-e^{-\eta }}{%
(1-e^{-\frac{1}{2}\eta })^{2}}$ as
\begin{eqnarray}
\nu ^{(\frac{1}{2})} &=&\frac{2}{1-\sqrt{\frac{\nu -1}{\nu +1}}}-1=\nu +%
\sqrt{\nu ^{2}-1}, \\
1-e^{-\eta } &=&1-\frac{\nu -1}{\nu +1}=\frac{2}{\nu +1}, \\
1-e^{-\frac{1}{2}\eta } &=&1-\sqrt{\frac{\nu -1}{\nu +1}}=\frac{\sqrt{\nu +1}%
-\sqrt{\nu -1}}{\sqrt{\nu +1}}, \\
(1-e^{-\frac{1}{2}\eta })^{2} &=&2\frac{\nu -\sqrt{\nu ^{2}-1}}{\nu +1}, \\
\frac{1-e^{-\eta }}{(1-e^{-\frac{1}{2}\eta })^{2}} &=&\frac{1}{\nu -\sqrt{%
\nu ^{2}-1}}=\nu +\sqrt{\nu ^{2}-1}=\nu ^{(\frac{1}{2})}. 
\end{eqnarray}

From Eqs. (\ref{eq4-14},\ref{eq4-21}), we have
\begin{eqnarray}
&&V^{(\frac{1}{2})}+OV^{(\frac{1}{2})}O 
=2\nu ^{(\frac{1}{2})}\left(
\begin{array}{cc}
\cosh (2|\zeta |)+\cos \theta \sinh (2|\zeta |) & 0 \\
0 & \cosh (2|\zeta |)-\cos \theta \sinh (2|\zeta |)%
\end{array}%
\right) , \\
&&\det [V^{(\frac{1}{2})}+OV^{(\frac{1}{2})}O] 
=[2\nu ^{(\frac{1}{2})}]^{2}[\cosh ^{2}(2|\zeta |)-\cos ^{2}\theta \sinh
^{2}(2|\zeta |)]    \nonumber \\
&&\ \ \ \ \ \ \ \ \ \ \ \ \ \ \ \ \ \ \ \ \ \ \ \ \ \ \ \  =[2\nu ^{(\frac{1}{2})}]^{2}[1+\sinh ^{2}(2|\zeta |)-\cos ^{2}\theta \sinh
^{2}(2|\zeta |)] 
=[2\nu ^{(\frac{1}{2})}]^{2}[1+\sin ^{2}\theta \sinh ^{2}(2|\zeta |)], \\
&&\lbrack V^{(\frac{1}{2})}+OV^{(\frac{1}{2})}O]^{-1} 
=\frac{1}{2\nu ^{(\frac{1}{2})}}\left(
\begin{array}{cc}
\frac{1}{\cosh (2|\zeta |)+\cos \theta \sinh (2|\zeta |)} & 0 \\
0 & \frac{1}{\cosh (2|\zeta |)-\cos \theta \sinh (2|\zeta |)}%
\end{array}%
\right) , \\
&&\exp \{-\frac{1}{2}(\overline{X}-\overline{X}^{\prime })^{T}[V^{(\frac{1}{2%
})}+OV^{(\frac{1}{2})}O]^{-1}(\overline{X}-\overline{X}^{\prime })\} 
=\exp \{-\frac{\overline{X}_{2}^{2}}{\nu ^{(\frac{1}{2})}[\cosh (2|\zeta
|)-\cos \theta \sinh (2|\zeta |)]}\}.
\end{eqnarray}

Taking these results into Eq. (\ref{eq4-29}), Eq. (\ref{eq4-32}) then follows.
\end{widetext}

%

\end{document}